\documentclass[12pt]{iopart}

\pdfoutput=1
\usepackage{amsfonts}
\usepackage{graphicx}
\usepackage{hyperref}
\usepackage{color}
\usepackage{subfigure}
\usepackage{textcomp}
\usepackage{cite}
\usepackage{float}
\usepackage{soul}
\usepackage{stackrel}
\usepackage{tikz}
\usepackage[mathscr]{eucal}

\definecolor{dgreen}{rgb}{0,0.7,0}

\expandafter\let\csname equation*\endcsname\relax
\expandafter\let\csname endequation*\endcsname\relax
\usepackage{amsmath,amssymb}

\newenvironment{rcases}
  {\left.\begin{aligned}}
  {\end{aligned}\right\rbrace}

\pdfoutput=1
\usepackage{esint}

\usepackage{color}
\definecolor{dgreen}{rgb}{0,0.7,0}

\begin{document}

\title{Extremal statistics of a one dimensional run and tumble particle with an absorbing wall}

\author{Prashant Singh, Saikat Santra and Anupam Kundu}

\address{International Centre for Theoretical Sciences, TIFR, Bengaluru 560089, India}
\ead{prashant.singh@icts.res.in}
\vspace{10pt}
\date{\today}
\begin{abstract}
\noindent
We study the extreme value statistics of a run and tumble particle (RTP) in one dimension till its first passage to the origin starting from the position $x_0~(>0)$. This model has recently drawn a lot of interest due to its biological application in modelling the motion of certain species of bacteria. Herein, we analytically study  the exact time-dependent propagators for a single RTP in a finite interval with absorbing conditions at its two ends. By exploiting a path decomposition technique, we use these propagators appropriately to compute  the joint distribution $\mathscr{P}(M,t_m)$ of the maximum displacement $M$ till first-passage and the time $t_m$ at which this maximum is achieved exactly.  The  corresponding marginal distributions $\mathbb{P}_M(M)$ and $P_M(t_m)$ are studied separately and verified numerically. In particular, we find that the marginal distribution $P_M(t_m)$ has interesting asymptotic forms for large and small $t_m$. While  for small $t_m$, the distribution $P_M(t_m)$ depends sensitively on the initial velocity direction $\sigma _i$ and is completely different from the Brownian motion, the large $t_m$ decay of $P_M(t_m)$ is same as that of the Brownian motion although the amplitude crucially depends on the initial conditions $x_0$ and $\sigma _i$. We verify all our analytical results to high precision by numerical simulations.

  
\end{abstract}

\section{Introduction}
While the mean and the variance tell us about the typical behaviour of a system, understanding extreme fluctuations is also important as they can have catastrophic consequences. For instance, from natural calamities like earthquake, tsunamis and floods to economic collapses and outbreak of pandemic are all examples of extreme events which can lead to devastating consequences {\cite{EVS-review-1,EVS-review-2,EVS-review-3,EVS-review-4,EVS-review-5,EVS-review-6,EVS-review-7,EVS-review-cor}}. In probability theory, the branch of extreme-value statistics deals with the study of the extreme deviations of a random process from its mean behaviour. Gnedenko's classical law of extremes tells us that for independent and identically distributed $N$ random variables, the distribution of the maximum (or minimum) possesses a scaling behaviour for large $N$ and the corresponding scaling function only depends on the asymptotic behaviour of the parent distribution from which these variables are drawn {\cite{EVS-review-3}}. Since then, there has been a plethora of studies to understand the extreme value for many weakly and strongly correlated processes \cite{Redner, disorder-1, KPZ-1, KPZ-2, RM-1, RM-3, EVS-correlated-2, EVS-pss-1, EVS-pss-2, EVS-pss-3, EVS-correlated-4, EVS-con-3,EVS-con-4, EVS-con-5, EVS-con-6}. We refer the readers to \cite{EVS-review-5} {and} \cite{EVS-review-cor} for a review on the extreme-value statistics.

While the knowledge of the value of the maximum $M$ is important, the time $T_m$ at which this maximum occurs is also important \cite{EVS-pss-1, Levy, Andersen}. Paradigmatic example is the one dimensional Brownian walker of fixed duration $t$ for which P. L\'{e}vy showed that the distribution of {$T _m$} is given by {\cite{Levy}}
\begin{align}
P _{BM}\left(T_m|t \right) = \frac{1}{\pi \sqrt{T_m \left( t-T_m \right)}},~~~~~\text{with }0 \leq T_m \leq t.
\label{levy-arc}
\end{align}
This distribution has square-root divergences both at $T_m \to 0$ and $T_m \to t$. On the other hand, it has the minimum value at $T_ m = t /2$ which is also the average of $T_m$. This implies that the typical value of $T_m$ is $0$ or $t$ while its average value is $t/2$. Over the years, the statistical properties of $T_m$ have been studied for a wide range of random processes like constrained and confined Brownian motion {\cite{tmax-2, confined-tmax}}, fractional Brownian motion {\cite{tmax-FBM-1, tmax-FBM-2, tmax-FBM-3}}, Black-Scholes model {\cite{tmax-3}}, random acceleration \cite{tmax-RAP}, continuous and discrete time random walks and vicious walks {\cite{tmax-CTRW, EVS-pss-1, tmax-vicious}}, active models {\cite{tmax-RTP-1, Mori2020}}, resetting systems {\cite{reset-tmax}}, fluctuating interface \cite{tmax-interface-growth} and heterogeneous diffusion processes {\cite{Heterogeneous-tmax}}. Extension to study the distribution of the time difference between the minimum and the maximum for stochastic processes has also been made in \cite{tdiff-1, tdiff-2}. 
Quite remarkably, the statistics of $T_m$ has found applications in convex hull problems \cite{convex} and also in detecting whether a stationary process is equilibrium or not \cite{confined-tmax}.

{While $M$ and $T_m$ for a process of a fixed time duration is important, the study of these quantities for a process until a stopping time (e.g., the first passage time to some prefixed value) is also important in several contexts. In the context of queueing theory, assuming the queue length performs a random walk, these two quantities would correspond to the maximum queue length and the time at which this length is achieved before the queue length gets to zero \cite{tmax-1}. Similarly in stock market, $x(t)$ represents the price of a stock that is often modelled by an exponential of the Brownian motion \cite{tmax-3}. Usually an agent can hold the stock till its price reaches a certain threshold value (say $x_{\rm th}$) since retaining the stock further can be too risky. Then the stopping time is simply the time at which the price falls below $x_{\rm th}$. Under this circumstance, obtaining the statistics of the time taken to achieve the maximum price is clearly important as this gives an idea about the best time to sell the stock. Another example arises in the biological context where knowledge about the maximal excursions of the tracer proteins before binding at a site is important \cite{Benichou, Pal2019}. While a long excursion is detrimental to the chemical reactions taking place locally, it might be useful if a large lifetime of the protein is favoured. A simple measure of such excursions is just to look at the maximum displacement achieved by the tracer. Hence, it is important to study the statistics of the maximum displacement till a stopping time and the time at which this maximum is attained.}
%

For the simple case of Brownian motion in one dimension, the marginal distributions of $M$ and $t_m$  till the first passage time to the origin, starting from some position $x_0~(>0)$, have been studied and they are given by {\cite{tmax-1}}
\begin{align}
\mathbb{P}_M^{BM}(M|x_0) = \frac{x_0}{M^2},~~~~ P_M^{BM}(t_m|x_0) = \frac{1}{2 \pi t_m}~\left[ \pi - \int _0 ^{\pi}dy~\vartheta \left( \frac{y}{2}, e^{-\frac{y^2 D t_m}{x_0^2}}    \right)  \right], \label{BM-results}
\end{align}
where $D$ is the diffusion constant and $\vartheta(y,z)$ denotes the fourth of Jacobi's theta functions. Moreover, the distribution $P_M^{BM}(t_m|x_0)$ was shown to have power-law forms at both large and small tails with $P_M^{BM}(t_m|x_0) \sim t_m^{-1/2}$ as $t_m \to 0$ and $P_M^{BM}(t_m|x_0) \sim t_m^{-3/2}$ as $t_m \to \infty$. However, to the best of our knowledge, the distribution of $t_m$ has not been calculated for any other stochastic process. In this paper we study the distribution of  $M$ and $t_m$ in the context of a {run and tumble particle in one dimension which has recently drawn a lot of interest.} {\cite{Berg2003}}. 

Previously known in the literature as persistent random walk \cite{RTP-1, RTP-2, RTP-3, RTP-4, RTP-5}, the run and tumble model has recently garnered significant interest due to its biological application in modelling the motion of certain species of bacteria like E. Coli \cite{Berg2003}. In this model, the position of the particle evolves as
{ 
\begin{align}
\frac{dx}{d t} = v ~\sigma (t), \label{RTP-motion}
\end{align}
}
where $v$ is the speed of the particle and $\sigma(t)$ is the dichotomous noise that can alternate randomly between two values $\pm 1$ at some rate $\gamma $. The values of noise at two different times are exponentially correlated as $v^2\langle \sigma (t_1)~\sigma (t_2) \rangle =v^2 \text{exp} \left( -2 \gamma |t_1 - t_2 |\right)$, which makes $x(t)$ in Eq. \eqref{RTP-motion} a non-Markov process. However in the limit $v \to \infty$ and $\gamma \to \infty$ with $D = v^2 / 2 \gamma$ fixed, the process reduces to the Brownian motion with effective diffusion constant $D$. On the other hand, for $\gamma \to 0 $ and $v \neq 0$, the noise $\sigma(t)$ in Eq. \eqref{RTP-motion} is just a constant and then the run and tumble particle (RTP) simply moves ballistically. Hence, one can interpolate between ballistic motion and diffusive motion by changing the parameters $v$ and $\gamma$. At finite values of these parameters, the {RTP model breaks the detailed balance condition and therefore has been an emblematic model for the active particles \cite{Cates2008}.} 

{Active matter refers to a general class of non-equilibrium systems whose microscopic constituents break the detailed balance condition. Consuming energy from the surroundings at microscopic scale, these particles can generate an autonomous motion via some internal mechanisms \cite{active_review1,active_review2}.}  Many studies have shown that {active particles} exhibit rich and distinctly non-thermal behaviours like motility induced phase separation \cite{Cates2015}, lack of equation of state in pressure {\cite{pressure-bas}}, flocking {\cite{Flocks}}, jamming \cite{Slowman2016} and so on. Moreover, many interesting features have been observed even for the non-interacting RTPs. Examples include - non-trivial large deviations and condensation transitions for the free space position distributions {\cite{dist-1,dist-4,dist-5,dist-6,dist-7,dist-8, new-dist-1}}, non-Boltzmann steady state distributions {\cite{dist-2, dist-3, entropy-1, entropy-2}}, universal first-passage properties and extremal statistics {\cite{FPT-1, FPT-3, Mori2020, tmax-RTP-1, FPT-5, FPT-6, FPT-7}}, functional statistics \cite{local}, convex hull problems \cite{convex-1, convex-2}, tagged particle properties {\cite{TP-1, TP-2, TP-3, TP-4}} and so on.

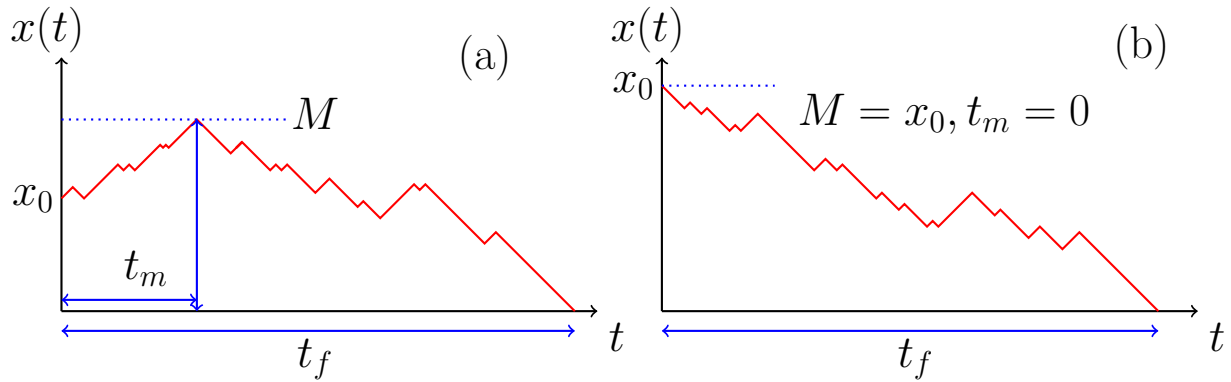
\begin{figure}[t]
	\begin{minipage}[h]{0.3\textwidth}
		\begin{tikzpicture}[scale=0.75]
		\draw[thick,->] (0,0) -- (9.5,0)node[anchor=north west] {\Large $t$};
		
		\node (c) at (-0.25,5) {\Large $x(t)$};
		
		\draw[thick,->] (0,0) -- (0,4.5); 
		
		\draw[thick,red] (0,2) -- (0.2,2.2) -- (0.4,2.0) -- (1.0,2.6)-- (1.1,2.5)-- (1.2,2.6) -- (1.3,2.5)-- (1.4,2.6)-- (1.75,2.95)-- (1.8,2.9)--(1.85,2.95)--(1.9,2.9)--(2.4,3.4)-- (2.8,3.0)-- (3.0,2.8)-- (3.2,3.0)-- (3.1,2.9) -- (3.2,3.0) -- (3.7,2.5) -- (3.8,2.6) -- (3.9,2.5) -- (4.0,2.6) -- (4.5,2.1) -- (4.75,2.35) -- (5.25,1.85) -- (5.35,1.95) -- (5.65,1.65)-- (6.25,2.25)-- (6.35,2.15)--(6.45,2.25)--(7.5,1.2)--(7.7,1.4)--(9.1,0.0) ;
		
		\draw[thick,<->,blue] (2.4,0) -- (2.4,3.4);
		
		\draw[thick,<->, blue] (0,0.2) -- (2.4,0.2);
		
		\draw[thick,<->,blue] (0,-0.35) -- (9.1,-0.35);
		
		\draw[thick,dotted, blue] (0,3.4) -- (4,3.4);
		
		\draw (4.5,-0.85) node {\Large $t_f$};
		
		\draw (1.5,0.75) node {\Large $t_m$};
		
		\draw (4.5,3.5) node {\Large $M$};
		
		\draw (-0.5,2) node {\Large $x_0$};
		
		\draw (7.5,4.5) node {\Large (a)};
		
		\end{tikzpicture}
	\end{minipage}
	\hspace*{3cm}
	\begin{minipage}[h]{0.1\textwidth}
		\begin{tikzpicture}[scale=0.750]
		\draw[thick,->] (0,0) -- (9.5,0)node[anchor=north west] {\Large $t$};
		\draw[thick,->] (0,0) -- (0,4.5); 
		\node (c) at (-0.25,5) {\Large $x(t)$}; 
		
		\draw[thick,red] (0,4) -- (0.4,3.6) -- (0.5,3.7) -- (0.7,3.5)-- (0.8,3.6)-- (1.2,3.2) -- (1.3,3.3)-- (1.4,3.2)-- (1.7,3.5)-- (2.7,2.5)--(2.9,2.7)--(3.1,2.5)--(3.2,2.6)-- (3.8,2.0)--(3.9,2.1)--(4.0,2.0)-- (4.2,1.8) -- (4.3,1.9) -- (4.7,1.5)-- (4.8,1.6)-- (4.9,1.5) -- (5.0,1.6)-- (5.5,2.1)-- (5.9,1.7)-- (6.0,1.8)--(6.1,1.7)--(6.5,1.3)--(6.7,1.5)-- (7.1,1.1)-- (7.2,1.2) --(7.4,1.4) --(8.8,0.0);
		\draw[dotted, thick, blue] (0,4) -- (2,4);
		
		\draw[thick,<->,blue] (0,-0.35) -- (8.8,-0.35);
		
		\draw (4.5,-0.85) node {\Large $t_f$};
		
		\draw (5.0,3.5) node {\Large $M=x_0,t_m=0$};
		
		\draw (-0.5,4) node {\Large $x_0$};
		
		\draw (8.5,4.7) node {\Large (b)};
		
		\end{tikzpicture}
	\end{minipage}
	\caption{(a) Schematic diagram of a typical trajectory of a RTP (shown in red) that starts from the position $x_0~(>0)$ and is observed till its first-passage time $t_f$ to the origin. Here $M$ is the maximum displacement of the particle before time $t_f$ and this displacement is achieved at time $t_m$. (b) An illustration of a trajectory (in red) where the RTP, starting from $x_0$ with negative velocity, does not cross $x=x_0$ till time $t_f$. Clearly for such trajectories $M=x_0$ and $t_m = 0$.}
	\label{Schematic}
\end{figure}

In this paper, we study the statistical properties of $M$ and $t_m$ (discussed above) for a one dimensional RTP. {We have schematically illustrated these quantities in Fig. \ref{Schematic}.} Our goal is to demonstrate how the persistent nature of the RTPs affects the statistics of these quantities. To this end, we use the path decomposition technique for Markov processes to write the joint distribution of $M$ and $t_m$ completely in terms of the propagators in a finite interval $[0,M]$ with absorbing boundary conditions at $x=0$ and $x=M$ and the associated exit probabilities from the origin without crossing $x=M$ {\cite{tmax-2, tmax-1}}. This means that the original problem of computing the extremal statistics reduces to the calculation of these propagators with two absorbing walls. Following \cite{RTP-3}, we calculate the exact time-dependent propagators of a RTP in a finite interval which are then used to compute the joint distribution. Finally, we marginalise this joint distribution appropriately to obtain the distributions of $M$ and $t_m$, pointing out the key differences from the passive Brownian motion.

The rest of our paper is structured as follows: In Sec. \ref{prob-abs-sec}, we compute the exact time-dependent propagators of a RTP in a finite interval $[0,M]$ with absorbing boundary condition at its two ends. These propagators were previously computed in \cite{RTP-3} in the context of telegrapher's equation. We revisit this computation here for a run and tumble particle for two reasons. Firstly, we reveal several detailed features of the propagators and compare them with rigorous numerical simulations. In the RTP language, we find interesting physical interpretations of these features as well as we  discuss various limits such as -- free RTP limit and free BM limit. Secondly, these expressions of the propagators become instrumental in addressing the extremal statistics till the first-passage time. We, therefore, re-derive these propagators in Sec. \ref{prob-abs-sec} which are then used to compute the exit probabilities in Sec. \ref{exit}. Sec. \ref{joint_dist} is devoted to the calculation of the joint distribution of $M$ and $t_m$. We compute the marginal distribution of $M$ in Sec. \ref{sec-prob-max} and that of $t_m$ in Sec. \ref{sec-prob-tm}. Finally, we summarise in Sec. \ref{conclusion}.  

\section{Probability distribution with two absorbing walls at $x=0$ and $x=M$}
\label{prob-abs-sec}
We begin with the computation of the probability distribution $P_{\sigma _f}(x,t|x_0, \sigma _i )$ of finding the RTP at position $x$ at time $t$ with velocity direction $\sigma _f$ in presence of two absorbing walls at $x= 0$ and $x=M~(>0)$ such that the particle was initially at $x_0~(0 \leq x_0 \leq M)$ with direction $\sigma _i$. Both $\sigma _i$ and $\sigma _f$ can take values $+1$ and $-1$. The propagators $P_{\pm}(x,t|x_0, \sigma _i )$ satisfy the Telegrapher's equations \cite{dist-1}
\begin{align}
\begin{split}
  \partial_t P_+(x,t|x_0, \sigma _i) &= -v \partial_x P_+(x,t|x_0, \sigma _i) -\gamma P_+(x,t|x_0, \sigma _i) + \gamma P_-(x,t|x_0, \sigma _i), \\
   \partial_t P_-(x,t|x_0, \sigma _i) &=~~~v \partial_x P_-(x,t|x_0, \sigma _i) +\gamma P_+(x,t|x_0, \sigma _i) - \gamma P _-(x,t|x_0, \sigma _i).
 \end{split}
\label{fokker}
\end{align}
To solve this set of coupled differential equations, we need to specify the initial conditions and the boundary conditions. Since the particle is initially located at $x_0$ with direction $\sigma _i$, the  initial condition is written as 
\begin{align}
P_{\pm}(x,0|x_0, \sigma _i) = \delta _{\sigma _i,\pm 1}~\delta (x-x_0).
\label{IC}
\end{align}
On the other hand, the boundary conditions are
\begin{align}
& P_+(x \to 0,t|x_0, \sigma _i)= 0, \label{bc1}~\\
& P_-(x \to M,t|x_0, \sigma _i) = 0 \label{bc2}.
\end{align}
For the particle to reach the origin with positive velocity, it must have crossed the origin at some earlier time before since the initial position $x_0 >0$. But the particle gets absorbed instantly whenever it crosses the origin. Hence, it can never reach the origin with positive velocity. This gives rise to the first boundary condition in Eq. \eqref{bc1}. Same physical reasoning gives the second boundary condition in Eq. \eqref{bc2} since the particle can never reach the wall at $x=M$ with negative velocity.

\noindent
To solve the master equations \eqref{fokker}, we perform the transformation
\begin{align}
P_{\pm}(x,t|x_0, \sigma _i)=e^{-\gamma t}~\mathcal{P}_{\pm}(x,t|x_0, \sigma _i), \label{ptansform}
\end{align}
and rewrite Eqs. \eqref{fokker} in terms of these new variables as
\begin{align}
\begin{split}
  \partial_t \mathcal{P}_+(x,t|x_0, \sigma _i) &= -v \partial_x \mathcal{P}_+(x,t|x_0, \sigma _i) + \gamma \mathcal{P}_-(x,t|x_0, \sigma _i), \\
   \partial_t \mathcal{P}_-(x,t|x_0, \sigma _i) &=~~~v \partial_x \mathcal{P}_-(x,t|x_0, \sigma _i) +\gamma \mathcal{P}_+(x,t|x_0, \sigma _i) .
 \end{split}
\label{fokkertran}
\end{align}
Observe that the initial and the boundary conditions do not change under these transformations. For simplicity, we also assume that the initial velocity direction is positive, i.e. $\sigma _i = +1$. The other case of $\sigma _i = -1$ will be considered later. To solve Eqs. \eqref{fokkertran}, we take the Laplace transformation with respect to $t~(\to s)$ as 
\begin{align}
\bar{\mathcal{P}}_{\pm}(x,s|x_0, +1)=\int_{0}^{\infty} dt~ e^{-s t}~\mathcal{P}_{\pm}(x,t|x_0, +1)
\end{align}
and use it in Eqs. \eqref{fokkertran} to obtain
\begin{align}
\begin{split}
\left[~v \partial_x+s~\right]\bar{\mathcal{P}}_{+}(x,s|x_0, +1)&=\gamma \bar{\mathcal{P}}_-(x,s|x_0, +1)+ \delta (x-x_0),\\
\left[-v \partial_x+s\right]\bar{\mathcal{P}}_{-}(x,s|x_0, +1)&=\gamma \bar{\mathcal{P}}_+(x,s|x_0, +1) .
\end{split}
\label{LT-FP-2}
\end{align}
\begin{figure}[t]
\includegraphics[scale=1.15]{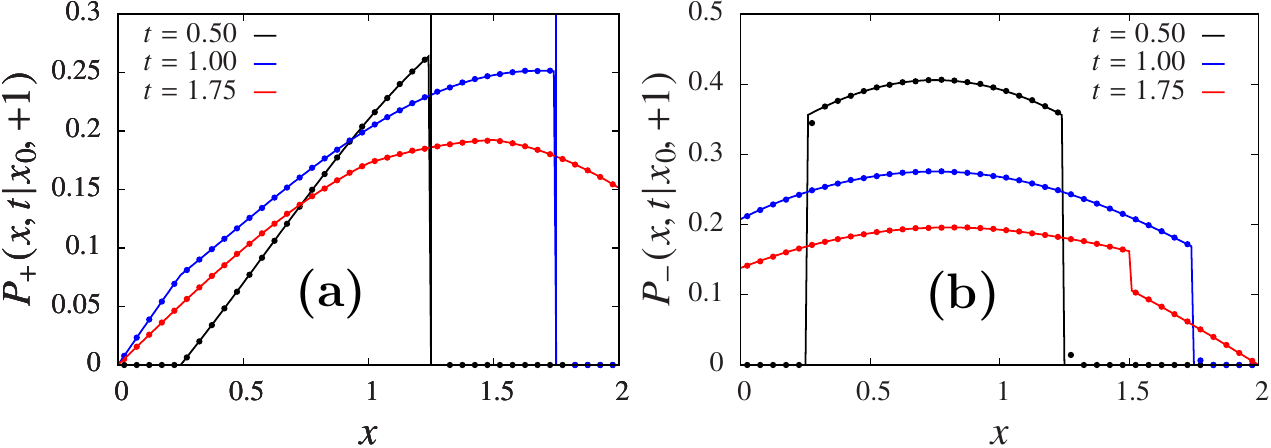}
\centering
\caption{Numerical verification of the propagators $P_{\pm}(x,t|x_0, +1)$ for $x_0=0.75$, $M=2$, $\gamma =1.5$, $ v=1$ and three different values of $t$. The solid lines represent the analytic expressions in Eqs. \eqref{probppp} and \eqref{probppm} while the symbols are the simulation data.}    
\label{probpic1}
\end{figure}
\begin{figure}[t]
\includegraphics[scale=1.15]{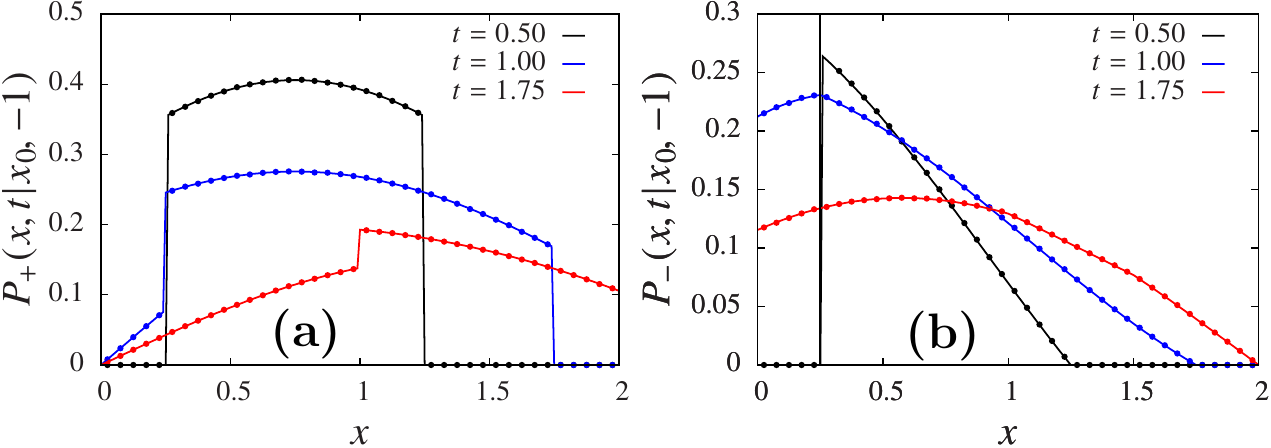}
\centering
\caption{Numerical verification of the propagators $P_{\pm}(x,t|x_0, -1)$ for $x_0=0.75$, $M=2$, $\gamma =1.5$ and $ v=1$ and three different values of $t$. The solid lines represent the analytic expressions in Eqs. \eqref{probpmm} and \eqref{probpmp} while the symbols are the simulation data.}    
\label{probpic2}
\end{figure}
In \ref{propagator_abs}, we have solved these two equations explicitly and get the Laplace transforms as
\begin{align}
\bar{\mathcal{P}}_+(x,s|x_0,+1)=&
    \frac{s+\beta \lambda(s)}{2 v \lambda(s)}\Big[ \left\{ \mathcal{F}(s) \right\}^{1+\beta}\mathcal{Z}(2 M-|x_0-x|,s)-\left\{ \mathcal{F}(s) \right\}^{\frac{1+\beta}{2}}\mathcal{Z}(2 M-x_0-x,s) \nonumber \\
 & ~~~~~~~~~~~+\mathcal{Z}(|x_0-x|,s)-\left\{ \mathcal{F}(s) \right\}^{\frac{1+\beta}{2}}\mathcal{Z}(x_0+x,s) \Big],\label{prob-LT-1}\\
  \bar{\mathcal{P}}_-(x,s|x_0,+1)=&\frac{\gamma}{2 v \lambda(s)}\Big[\mathcal{F}(s) \mathcal{Z}(2 M-|x_0-x|,s)-\mathcal{Z}(2 M-x_0-x,s) \nonumber \\
  & ~~~~~~~~~~~+\mathcal{Z}(|x_0-x|,s)-\mathcal{F}(s)\mathcal{Z}(x_0+x,s) \Big].
\label{prob-LT-2}
\end{align}
Here $\lambda(s) = \sqrt{s^2 -\gamma ^2}$ and $\beta  = \text{sgn}(x-x_0)$ is $1$ if $x >x_0$ and $-1$ otherwise. Also the functions $\mathcal{F}(s)$ and $\mathcal{Z}(x,s)$ are given in Eqs. \eqref{abs_appen10}. One now has to perform the inverse Laplace transformation to get the distributions in the time domain. Using the inversions in Eqs. \eqref{abs_appen12} and \eqref{abs_appen13}, we find
\begin{align}
P_{+}(x,t|x_0,+1)&=e^{-\gamma t} \delta(x-x_0-vt)+\frac{e^{-\gamma t}}{2 v}\Big[\mathcal{U}_{\beta}^{0}(\mid x-x_0\mid,t)-\mathcal{U}_{\beta}^{\frac{1+\beta}{2}}( x+x_0,t) \nonumber\\
&~~~~~- \mathcal{U}_{\beta}^{\frac{1+\beta}{2}}( 2M-x-x_0,t)+\mathcal{U}_{\beta}^{1+\beta}( 2M-\mid x-x_0\mid,t)\Big]\label{probppp}\\
P_{-}(x,t \mid x_0,+1)&=\frac{\gamma e^{-\gamma t}}{2 v}\Big[ \mathcal{W}_0(\mid x-x_0 \mid,t)-\mathcal{W}_0(2M- x-x_0 ,t)-\mathcal{W}_1(x+x_0,t) \nonumber\\
&~~~~~~~~~~~~~~~~~~~+\mathcal{W}_1(2M-\mid x-x_0 \mid,t)\Big],\label{probppm}
\end{align}
where the functions $\mathcal{U}_{\beta}^{(m)}(x,t)$ and $\mathcal{W}_{m}(x,t)$ are defined as 
\begin{align}
\mathcal{U}_{\beta}^{m}(x,t)=
& \sum_{n=0}^{\infty}\frac{\gamma}{2}\Big[g_{2n+2m-1}(x+2nM,t)\Big\{1+\beta\frac{x+2nM}{vt+x+2nM}\Big\} +g_{2n+2m+1}(x+2nM,t) \nonumber \\ 
&\times \Big\{ 1+\beta\frac{x+2nM}{vt-x-2nM}\Big\}
+\frac{4\beta t(n+m)}{\gamma t^2-\gamma\frac{(x+2nM)^2}{v^2}} g_{2n+2m}(x+2nM,t)\Big], \nonumber  \\&~~~~~~~~~~~~~~~~~~~~~~~~~~~~~~~~~~~~~~~~~~~~~~~~~~~~~~~~~~~~~~~~~~~~~~~~~\normalsize{\text{for }m \geq 1}, \label{U-fun-2}\\
\mathcal{U}_{\beta}^{0}(x,t) = & \sum_{n=0}^{\infty}\frac{\gamma}{2}\Big[g_{2n+1}(x+2(n+1)M,t)\Big\{1+\beta\frac{x+2(n+1)M}{vt+x+2(n+1)M}\Big\} \nonumber \\
&+g_{2n+3}(x+2(n+1)M,t)  \Big\{ 1+\beta\frac{x+2(n+1)M}{vt-x-2(n+1)M}\Big\} \nonumber \\
&+\frac{4\beta n t}{\gamma t^2-\gamma\frac{(x+2nM)^2}{v^2}} g_{2n}(x+2nM,t)\Big]+\gamma \frac{t+\beta \frac{x}{v}}{t-\frac{x}{v}}g_1(x,t), \label{U-fun} \\
\mathcal{W}_m(x,t) =& \sum_{n=0}^{\infty}g_{2n+2m}(x+2nM,t),~~~~~\text{for }m \geq 0, \label{W-fun}
\end{align}
with $g_m(x,t)$ given in terms of the modified Bessel function of the first kind $I_m(x,t)$ as
\begin{align}
g_m(x,t)= & \left(\frac{vt-x}{vt+x}\right)^{\frac{m}{2}} I_m\left(\frac{\gamma}{v}\sqrt{v^2t^2-x^2} \right)\theta \left(vt-x \right),~~~~~~~~ \text{for }m \geq 0. \label{g-fun}
\end{align}
 In this equation, $\theta(z)$ represents the Heaviside theta function which takes value $1$ if $z >0$ and $0$ otherwise. 

 Note that up to this point, we have assumed that the RTP initially starts from $x_0$ with positive velocity $(\sigma _i = +1)$. But one can carry out the same analysis for negative initial velocity $(\sigma _i = -1)$ also and get the exact form of the propagators $P_{\pm}(x,t|x_0,-1)$. However here, we instead use the symmetry of the problem which gives propagators with $\sigma _ i = -1$ completely in terms of the propagators with $\sigma _ i =+1$. It is easy to see that the problem has the following symmetry: 
\begin{align}
P_{\pm}(x,t|x_0,-1)=P_{\mp}(M-x,t \mid M-x_0,+1)
\label{relation_minusplus}.
\end{align}
Then inserting the expressions of $P_{\mp}(M-x,t \mid M-x_0,+1)$ from Eqs. \eqref{probppp} and \eqref{probppm} in Eq. \eqref{relation_minusplus}, we find the distributions $P_{\pm}(x,t|x_0,-1)$ to be
\begin{align}
P_{-}(x,t|x_0,-1)&=e^{-\gamma t} \delta(x-x_0+vt)+\frac{e^{-\gamma t}}{2 v}\Big[\mathcal{U}_{-\beta}^{0}(\mid x-x_0\mid,t)-\mathcal{U}_{-\beta}^{\frac{1-\beta}{2}}( x+x_0,t), \nonumber\\
&~~~~~- \mathcal{U}_{-\beta}^{\frac{1-\beta}{2}}( 2M-x-x_0,t)+\mathcal{U}_{-\beta}^{1-\beta}( 2M-\mid x-x_0\mid,t)\Big]\label{probpmm}\\
P_{+}(x,t \mid x_0,-1)&=\frac{\gamma e^{-\gamma t}}{2 v}\Big[ \mathcal{W}_0(\mid x-x_0 \mid,t)-\mathcal{W}_1(2M- x-x_0 ,t)-\mathcal{W}_0(x+x_0,t) \nonumber\\
&~~~~~~~~~~~~~~~~~~~+\mathcal{W}_1(2M-\mid x-x_0 \mid,t)\Big].\label{probpmp}
\end{align}
We have plotted the propagators $P_{\pm}(x,t|x_0, \pm 1)$ in Figs. \eqref{probpic1} and \eqref{probpic2} and have also compared them with the numerical simulations. Our analytic expressions are in excellent agreement with the numerical data. Compared to the Brownian motion {\cite{Redner}}, we see some key differences for the active case. Firstly, observe that the distributions $P_{\pm}(x,t|x_0, \pm 1)$ in Eqs. \eqref{probppp} and \eqref{probpmm} have delta-function terms like $e^{-\gamma t}~\delta(x-x_0 \mp vt)$. This contribution arises from those events for which the RTP has not changed its initial velocity direction till time $t$ and such events occur with probability $e^{-\gamma t}$. Also, for small times, the particle is yet to feel the presence of the absorbing walls. Therefore, we anticipate the propagators at small times to match with their expressions in the free space {\cite{dist-1}}. To see this, we use $I_m(z) \sim z^{m}$ as $z \to 0$ in Eq. \eqref{g-fun} which then gives $g_m(x,t) \sim (vt-x)^m$ as $t \to 0$. Inserting this form in the expressions of $\mathcal{U}_{\beta}^{m}(x,t)$ and $\mathcal{W}_{m}(x,t)$ in Eqs. \eqref{U-fun} and \eqref{W-fun} respectively, we see that the expressions for small $t$ become
\begin{align}
\mathcal{U}_{\beta}^{m}(x,t) \simeq & ~\delta _{m,0} ~\delta_{\beta,1}~\gamma \sqrt{\frac{vt+x}{vt-x}} I_1\left(\frac{\gamma}{v}\sqrt{v^2t^2-x^2} \right)\theta \left(vt-x \right), \label{small-U} \\
\mathcal{W}_{m}(x,t) \simeq &~\delta _{m,0}~I_0\left(\frac{\gamma}{v}\sqrt{v^2t^2-x^2} \right)\theta \left(vt-x \right).\label{small-W}
\end{align} 
Plugging these approximate forms in the expressions of the propagators in Eqs. \eqref{probppp}, \eqref{probppm}, \eqref{probpmm} and \eqref{probpmp} gives
\begin{align}
P_{\pm}(x,t|x_0,\pm 1) \simeq &  e^{-\gamma t}\delta(x-x_0 \mp vt)+\frac{\gamma~e^{-\gamma t}}{2v}\sqrt{\frac{vt\pm (x-x_0)}{vt \mp (x-x_0)}}~ \nonumber \\
& \times I_1\left(\frac{\gamma}{v}\sqrt{v^2t^2-(x-x_0)^2} \right)\theta \left(vt-|x-x_0| \right), \label{small-t-abs-1}\\
P_{\pm}(x,t|x_0,\mp 1) \simeq & ~\frac{\gamma~e^{-\gamma t}}{2v}~I_0\left(\frac{\gamma}{v}\sqrt{v^2t^2-(x-x_0)^2} \right)\theta \left(vt-|x-x_0| \right). \label{small-t-abs-2}
\end{align}
\begin{figure}[t]
	\includegraphics[scale=1.15]{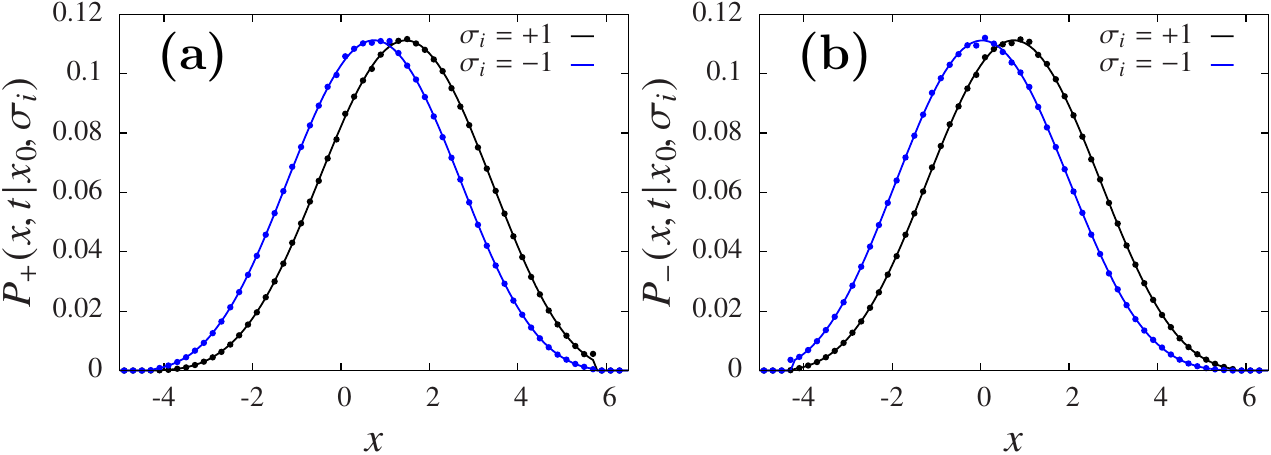}
	\centering
	\caption{Analytical expressions of the propagators $P_{\pm}(x,t|x_0, \sigma _i)$ for small times given in Eqs. \eqref{small-t-abs-1} and \eqref{small-t-abs-2} (shown by solid lines) have been plotted and compared with the numerical simulations of the free space position distributions (shown by filled circles). Parameters chosen are $x_0=0.75,~v=1,~\gamma=1.5~\text{and}~t=5$. For better clarity, we have not shown the delta functions in $P_{\pm}(x,t|x_0, \pm)$. }    
	\label{free_prob}
\end{figure}
We have numerically verified these small-$t$ forms in Fig. \ref{free_prob} which also converge with the free space position distributions given in \cite{dist-1}. However, note that these expressions are correct only at small times and become invalid once the particle hits the absorbing walls. Then one needs to consider the full expression given before. Interestingly, in Figs. \ref{probpic1}(b) and \ref{probpic2}(b), we see that distributions $P_{\pm}(x,t|x_0, \mp 1)$ possess discontinuity at some $x^{\ast}$ for a given $t$ [in Fig.~\ref{probpic1} (b), $x^{\ast}=1.5$ for $t = 1.75$]. To understand this physically, let us focus on  $P_{+}(x,t|x_0, -1)$ in Fig. \ref{probpic1}(b). For this case, the RTP will atleast take time $t_M = \frac{M-x_0}{v} $ to reach the absorbing wall at $x=M$. Recall that it can reach $x=M$ only with positive velocity direction. In the subsequent time interval $[t_M,t_M+dt]$, some RTPs will get absorbed by the wall while the remaining fraction will flip their direction from $+v$ to $-v$ for any non-zero $\gamma$. This remaining fraction will travel a distance $v(t-t_M)$ in the time interval $(t-t_M)$ and reach the position $x^{\ast} = M -v(t-t_M) = (2M-vt-x_0 )$. Therefore, we get a discontinuity in the distribution $P_{+}(x,t|x_0, -1)$ at position $x^{\ast}$. For instance, for $t = 1.75$ and given parameters in Fig. \ref{probpic1}(b), we find $x^{\ast} = 1.5$. By virtue of the symmetry $P_{+}(x,t|x_0,-1)=P_{-}(M-x,t \mid M-x_0,+1)$ discussed above, we also get a discontinuity at $x^{\ast} = (vt-x_0)$ for $P_{+}(x,t|x_0,-1)$.

Another contrasting feature for the active case is that the distributions $P_{\pm}(x,t|x_0, +1)$ in Eqs. \eqref{probppp} and \eqref{probppm} are non-zero even when $x_0 = 0$ whereas $P_{\pm}(x,t|x_0 \to 0^+, -1)$ in  Eqs. \eqref{probpmm} and \eqref{probpmp} are precisely zero. Similarly, when the initial position $x_0=M^-$, we see that $P_{\pm}(x,t|x_0 \to M^-, -1)$ are non-zero whereas $P_{\pm}(x,t|x_0 \to M^-, +1)$ are zero. For Brownian motion, the distribution is always zero for $x_0 = 0$ and $x_0=M$ \cite{tmax-1, Redner}. In order to understand this difference, let us look at the behaviour of $P_{\pm}(x,t|x_0, +1)$ when $x_0 = 0^+$. Due to the persistent nature of the active particle, it has a non-zero probability to survive the absorbing wall at the origin till time $t$ even if it starts from the origin but with positive velocity. This results in the non-vanishing distributions $P_{\pm}(x,t|x_0 \to 0^+, +1)$. On the other hand, the RTP will instantly get absorbed at the origin if it starts from the origin with negative velocity from $x_0 = 0$. Consequently, the distributions $P_{\pm}(x,t|x_0 \to 0^+, -1)$ also vanish for this case. Same physical arguments explain why $P_{\pm}(x,t|x_0, -1)$ for $x_0 = M^-$ are non-vanishing but $P_{\pm}(x,t|x_0 \to M^-, +1)$ vanish. 

\noindent
Finally, using the large-$z$ asymptotics of $I_m(z) \simeq e^{z}/\sqrt{2 \pi z}$, one can show that 
\begin{align}
& g_m(x,t)  \simeq \frac{1}{\sqrt{2 \pi \gamma t}}~\text{exp} \left( \gamma t -\frac{x^2}{4 Dt} \right),~~~\text{and} \\
& \mathcal{W}_m(x,t) ~\simeq ~\gamma^{-1} \mathcal{U}^m _{\beta}(x, t) ~\simeq~ \sum _{n=0}^{\infty}\frac{1}{\sqrt{2 \pi \gamma t}}~\text{exp} \left( \gamma t -\frac{(x+2nM)^2}{4 Dt} \right),
\end{align}
in the limit $\gamma \to \infty$ and $v \to \infty$ keeping the ratio $D=\frac{v^2}{2\gamma}$ fixed. 
Substituting these approximate forms in the expressions of the propagators in Eqs. \eqref{probppp}, \eqref{probppm}, \eqref{probpmm} and \eqref{probpmp}, we get
\begin{align}
P_{\sigma_f}(x,t|x_0, \sigma_i) \simeq & \frac{1}{2\sqrt{4 \pi D t}}\sum _{n=0}^{\infty}\Bigg[ \text{exp} \left(-\frac{(|x-x_0|+2nM)^2}{4Dt} \right)-\text{exp} \left(-\frac{(x+x_0+2nM)^2}{4Dt}\right) \Bigg. \nonumber\\
&\Bigg. -\text{exp} \left(-\frac{(2M-x-x_0+2nM)^2}{4Dt} \right) +\text{exp} \left(-\frac{(2M-|x-x_0|+2nM)^2}{4Dt}\right) \Bigg].\nonumber
\end{align}
Thus, the total propagator $P(x,t|x_0, \sigma_i)=P_{+}(x,t|x_0, \sigma_i)+P_{-}(x,t|x_0, \sigma_i)$ correctly reduces to the Brownian motion result for $\gamma \to \infty$ and $v \to \infty$ with fixed $D=v^2 / 2 \gamma $ \cite{Redner}. 
\section{Exit probability from an interval $[0,M]$}
\label{exit}
We now look at the RTP on a finite interval $[0,M]$ and calculate the probability $E_{\pm}(x_0, M)$ that the RTP starting from $x_0$ with velocity $\pm v$ exits from the boundary $x=0$ without crossing the boundary at $x=M$. As explained later, this quantity will be useful in computing the extremal statistics of a RTP observed till its first-passage time. Now to calculate these exit probabilities, one can, in principle, proceed in the same way as {\cite{dist-1, FPT-5}} where one writes the backward equations for $E_{\pm}(x_0, M)$ and solve them explicitly with appropriate boundary conditions. However, here we take a different route to calculate these probabilities.

Using the propagators derived in the previous section, one can write the flux $J_{\pm}(x_0,M,t)$ at the origin as
\begin{align}
J_{\pm}(x_0,M,t) = v \left[ P_{+}(0,t|x_0, \pm 1)+P_{-}(0,t|x_0, \pm 1) \right].
\label{new-exit-1}
\end{align}
Notice that the boundary condition in Eq. \eqref{bc1} gives the first term $P_{+}(0,t|x_0, \pm) = 0$. To write the second term, we use the transformation in Eq. \eqref{ptansform} and obtain
\begin{align}
J_{\pm}(x_0,M,t) = v e^{-\gamma t}  \mathcal{P}_{-}(0,t|x_0, \pm 1) .
\label{new-exit-2}
\end{align}
Noting that the exit probability is equal to the time integrated flux through the origin {\cite{Redner}}, we write
\begin{align}
E_{\pm}(x_0, M) & = \int _{0}^{\infty}dt~J_{\pm}(x_0,M,t)= v\bar{\mathcal{P}}_{-}(0,\gamma|x_0, \pm 1).
\end{align}
Recall that $\bar{\mathcal{P}}_{-}(0,s|x_0, \pm 1)$ denotes the Laplace transform of $\mathcal{P}_{-}(0,t|x_0, \pm 1)$ with respect to $t$. Now the exact form of $\bar{\mathcal{P}}_{-}(0,\gamma|x_0, +1)$ is given in Eq. \eqref{prob-LT-2} with $s \to \gamma$ and $x=0$. On the other hand, to compute $\bar{\mathcal{P}}_{-}(0,\gamma|x_0, -1)$ we reiterate the symmetry of the problem $\bar{\mathcal{P}}_{-}(0,\gamma|x_0, -1) = \bar{\mathcal{P}}_{+}(M,\gamma|M-x_0, +1) $ with $\bar{\mathcal{P}}_{+}(M,\gamma|M-x_0, +1)$ given in Eq. \eqref{prob-LT-1} with $s \to \gamma$ and $x=0$. Using these Laplace transforms, we find the expressions of $E_{\pm}(x_0, M)$ as
\begin{align}
&E_+(x_0,M)=\left(\frac{M-x_0}{M+\frac{v}{\gamma}}\right) \theta(M-x_0), \label{exit_eq6}\\
&E_-(x_0,M)=\left(1-\frac{x_0}{M+\frac{v}{\gamma}}\right) \theta (M-x_0). \label{exit_eq7}
\end{align}
Expectedly, in the limit $\gamma \to \infty$, $v \to \infty$ with $D = v^2 /2 \gamma $ fixed, we recover the result for the Brownian motion $E_{\pm}(x_0, M) = \left(1 -x_0 /M \right)$ {\cite{Redner, max-1}}.
\section{Joint distribution $\mathscr{P}(M,t_m|x_0, \pm 1)$ of $M$ and time $t_m$ before its first passage time $t_f$}
\label{joint_dist}
Let us now look at the joint distribution $\mathscr{P}(M,t_m|x_0, \pm 1)$ that the RTP, starting from $x_0$ with velocity $\pm v$, reaches its maximum displacement $M$ at time $t_m$ before getting absorbed at the origin. To compute this joint distribution, we decompose the full trajectory $\{ x(t);~0 \leq t \leq t_f \}$ into two parts: first part is $\{ x(t);~0 \leq t \leq t_m \}$  and the second part is $\{ x(t);~ t_m \leq t \leq t_f \}$. As illustrated in Fig. \ref{Schematic}(a), the RTP, in the first part, reaches the position $x=M$ at time $t_m$ for the first time without hitting the origin. Since $x_0 <M$, the particle can reach $x=M$ only with positive velocity. Therefore, the contribution of this part to the joint distribution $\mathscr{P}(M,t_m|x_0, \pm 1)$ is simply the propagator $P_{+}(M,t_m|x_0, \pm 1)$ with two absorbing walls at $x=0$ and $x=M$. Next in the second part, the RTP has to first undergo a tumble so that its velocity changes from $+v$ to $-v$ (since $M$ is the maximum displacement). Then, starting from $M$ with velocity $-v$, it will get absorbed at the origin without crossing the $x=M$ boundary again. Hence, the contribution of this part to $\mathscr{P}(M,t_m|x_0, \pm 1)$ is equal to $\gamma E_-(M,M)$, where a factor $\gamma$ comes due to a tumble that the particle experiences at time $t_m^+$. Since the run and tumble motion is a Markov process in $(x, \sigma)$ variables, the contributions associated to the two parts are statistically independent. This allows us to write the joint distributions $\mathscr{P}(M,t_m|x_0, \pm 1)$ as
\begin{align}
\mathscr{P}(M,t_m|x_0, \pm 1) = \gamma ~P_{+}(M,t_m|x_0, \pm 1)~E_-(M,M),
\label{new-JT-1}
\end{align}
where the functions $P_{+}(M,t_m|x_0, \pm 1)$ and $E_-(M,M)$ are given in Eqs. \eqref{probppp}, \eqref{probpmp} and \eqref{exit_eq7} as
\begin{align}
& P_{+}(M,t_m|x_0, +1) = \frac{e^{-\gamma t_m}}{2 v} \left[2v \delta(M-x_0-vt_m)    + \mathcal{U}_{1}^{0}(M-x_0,t_m)-\mathcal{U}_{1}^{1}(M-x_0,t_m) \right. \nonumber \\
&~~~~~~~~~~~~~~~~~~~~~~~~~~~~~~~~~~ \left. -~\mathcal{U}_{1}^{1}(M+x_0,t_m)+\mathcal{U}_{1}^{2}(M+x_0,t_m) \right], \label{pp1} \\
& P_{+}(M,t_m|x_0, -1) = \frac{\gamma e^{-\gamma t_m}}{2 v} \left[ \mathcal{W}_{0}(M-x_0,t_m)-\mathcal{W}_{1}(M-x_0,t_m) -~\mathcal{W}_{0}(M+x_0,t_m)\right. \nonumber \\
&~~~~~~~~~~~~~~~~~~~~~~~~~~~~~~~~~~~~~~~ \left. +\mathcal{W}_{1}(M+x_0,t_m) \right] \label{pp2}\\
& E_-(M,M)  = \frac{v}{\gamma \left(M+\frac{v}{\gamma} \right)}. \label{ee1}
\end{align}
Note that while writing $\mathscr{P}(M,t_m|x_0, \pm 1)$ in Eq. \eqref{new-JT-1}, we have assumed that the maximum $M$ is strictly greater than $x_0$. However, when the RTP starts with a negative velocity from $x_0$, we expect a non-zero contribution to the joint distribution $\mathscr{P}(M,t_m|x_0, -1)$ from those events where the particle starting from $x_0$ hits the origin without crossing $x=x_0$ again. Schematic illustration of such trajectories is shown in Fig. \ref{Schematic}(b). Clearly for such events $M=x_0$ and $t_m = 0$. The probability of observing such events is just the exit probability $E_-(x_0,x_0)$. Hence, the contribution of these trajectories to the joint distribution is
\begin{align}
\underbrace{\mathscr{P}(M,t_m|x_0, -1)}_{\text{contribution~from~paths~with~}M=x_0} = \delta(M-x_0) ~\delta (t_m) ~E_-(M,M).
\end{align}
Adding all the contributions, we find the final form of the joint distribution as
\begin{align}
& \mathscr{P}(M,t_m|x_0, +1) = \gamma ~P_{+}(M,t_m|x_0, +1)~E_-(M,M), \label{new-JT-2} \\
& \mathscr{P}(M,t_m|x_0, -1) = \delta(M-x_0) \delta (t_m) E_-(M,M)+ \gamma P_{+}(M,t_m|x_0, -1)E_-(M,M). \label{new-JT-3}
\end{align}
To summarise, Eqs.~\eqref{new-JT-2} and \eqref{new-JT-3} provide the joint distributions $\mathscr{P}(M,t_m|x_0, \pm 1)$ of the maximum displacement $M$ and time $t_m$ at which this maximum is attained for a one dimensional RTP evolved till its first-passage time $t_f$. In what follows, we will appropriately integrate this joint distribution to get the marginal distributions of $M$ and $t_m$ and analyze different asymptotic regimes.
\begin{figure}[t]
	\includegraphics[scale=1.15]{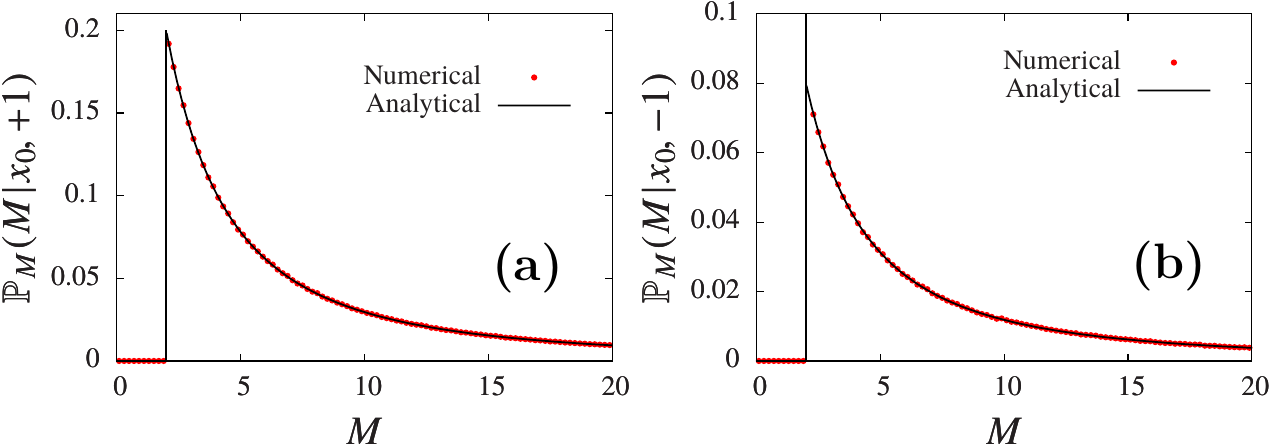}
	\centering
	\caption{Comparison of the analytical expressions of the distributions $\mathbb{P}_{M}(M|x_0, {\pm 1})$ given in Eqs. \eqref{PPmargM-eq-3} and \eqref{PPmargM-eq-4} with the numerical simulations for $x_0=2$, $\gamma =0.5$ and $v=1.5$. Notice that only $\mathbb{P}_{M}(M|x_0, -1)$ in panel (b) has a delta function peak at $M=x_0.$}    
	\label{PM-pic}
\end{figure}
\subsection{Marginal probability distribution of $M$}
\label{sec-prob-max}
We first look at the marginal distribution $\mathbb{P}_{M}(M|x_0, \pm 1) = \int _0 ^{\infty}dt_m~\mathscr{P}(M,t_m|x_0, \pm 1)$ of the maximum displacement $M$. Let us first consider $\mathbb{P}_{M}(M|x_0, + 1)$ for which we plug $\mathscr{P}(M,t_m|x_0, + 1)$ from Eq. \eqref{new-JT-2} and get
\begin{align}
\mathbb{P}_{M}(M|x_0, + 1) = \gamma E_-(M,M) \int _{0}^{\infty}dt_m ~P_{+}(M,t_m|x_0, + 1).
\label{PPmargM-eq-1}
\end{align}
Substituting the form of $P_{+}(M,t_m|x_0, +1)$ from Eq. \eqref{pp1} in this equation, we see that one then needs to compute the Laplace transform of different $\mathcal{U}$-functions that appear in Eq. \eqref{pp1}. Interestingly, these Laplace transforms can be computed exactly as shown in Eq. \eqref{abs_appen12}. Then the distribution $\mathbb{P}_{M}(M|x_0, +1)$ takes the form
\begin{align}
\mathbb{P}_{M}(M|x_0, +1) =   \frac{\text{exp} \left(-\frac{\lambda(\gamma)~(M-x_0)}{v} \right)-\mathcal{F}(\gamma)~\text{exp} \left(-\frac{\lambda(\gamma )~(M+x_0)}{v} \right)}{\left( M + \frac{v}{\gamma} \right)~\left[1-\mathcal{F}(\gamma)~\text{exp} \left(-\frac{2 \lambda(\gamma )M}{v} \right) \right]},
\label{PPmargM-eq-2}
\end{align}
with $\lambda(s) = \sqrt{s^2 - \gamma ^2}$ and function $\mathcal{F}(s)$ given in Eq. \eqref{abs_appen10}. From these expressions, it is clear that $\lambda\left(s \to \gamma \right) \simeq 0$ and $\mathcal{F}( s \to \gamma ) \simeq 1-\frac{2 \lambda\left(s \to \gamma \right)}{\gamma}$. With these approximations, the distribution $\mathbb{P}_{M}(M|x_0, +1)$ in Eq. \eqref{PPmargM-eq-2} becomes
\begin{align}
\mathbb{P}_{M}(M|x_0, +1) = \frac{x_0 + \frac{v}{\gamma}}{\left( M + \frac{v}{\gamma} \right)^2}~\theta(M-x_0). \label{PPmargM-eq-3}
\end{align}
Same analysis can be also carried out for the other distribution $\mathbb{P}_{M}(M|x_0, -1)$ when the RTP starts from $x_0$ with the negative velocity. For this case, we obtain
\begin{align}
\mathbb{P}_{M}(M|x_0, -1)= \frac{v}{v+\gamma M}~\delta (M-x_0)+\frac{x_0}{(M+\frac{v}{\gamma})^2}\theta(M-x_0). \label{PPmargM-eq-4}
\end{align}
In Fig. \ref{PM-pic}, we have compared our analytical expressions with the same obtained from the numerical simulations. We observe nice agreement between them. It is worth remarking that the exit probabilities $E_{\pm}(x_0,M')$ derived in Sec. \ref{exit} represent the cumulative probability that the maximum is less than $M '$, i.e. 
\begin{align}
E_{\pm}(x_0,M') = \text{Prob}[\text{max} \{ x(t)\} <M'].
\end{align}
Distributions $\mathbb{P}_{M}(M|x_0, {\pm 1})$ are then related to 
$E_{\pm}(x_0,M)$ as 
\begin{align}
\mathbb{P}_{M}(M|x_0, {\pm 1}) = \frac{\partial}{ \partial M}E_{\pm }(x_0,M).
\end{align}
It is easy to check that by inserting $E_{\pm}(x_0,M)$ from Eqs. \eqref{exit_eq6} and \eqref{exit_eq7}, we obtain exactly the same form of $\mathbb{P}_{M}(M|x_0, {\pm 1})$ as obtained in Eqs. \eqref{PPmargM-eq-3} and \eqref{PPmargM-eq-4}.
\subsection{Marginal probability distribution of $t_m$}
\label{sec-prob-tm}
We now look at the distributions $P_{M}(t_m|x_0, \pm 1)$ of time $t_m$ at which the maximum displacement $M$ is achieved before its first-passage time. Marginalising the joint distributions $\mathscr{P}(M,t_m|x_0, \pm 1)$ in Eqs. \eqref{new-JT-2} and \eqref{new-JT-3}, we find
\begin{align}
P_{M}(t_m|x_0, +1) & = \frac{v ~e^{-\gamma t_m}}{x_0 + vt_m + \frac{v}{\gamma}}+\frac{e^{-\gamma t_m}}{2} \int _{x_0}^{\infty} \frac{dM}{M+\frac{v}{\gamma}} \left[ \mathcal{U}_{1}^{0}(M-x_0,t_m)-\mathcal{U}_{1}^{1}(M-x_0,t_m) \right. \nonumber \\
&~~~~~~~~~~~~~~~~ \left. -~\mathcal{U}_{1}^{1}(M+x_0,t_m)+\mathcal{U}_{1}^{2}(M+x_0,t_m) \right], \label{Ptm-p} \\
P_{M}(t_m|x_0, -1) & = \frac{v~\delta(t_m)}{\gamma x_0+v}+\frac{\gamma e^{-\gamma t_m}}{2} \int _{x_0}^{\infty} \frac{dM}{M+\frac{v}{\gamma}} \left[ \mathcal{W}_{0}(M-x_0,t_m)-\mathcal{W}_{1}(M-x_0,t_m) \right. \nonumber \\
&~~~~~~~~~~~~~~~~ \left. -~\mathcal{W}_{0}(M+x_0,t_m) +\mathcal{W}_{1}(M+x_0,t_m) \right], \label{Ptm-n}
\end{align}
where the $\mathcal{U}$ and $\mathcal{W}$ functions are given in Eqs. \eqref{U-fun} and \eqref{W-fun} respectively. Performing analytically the integration over $M$ in these equations turns out to be difficult. However, one can make some progress for different limits of $t_m$. For example looking at the expressions of $\mathcal{U}$ and $\mathcal{W}$ functions, we see that only few terms in the series contribute for $t_m \to 0$ due to the presence of various $\theta$-functions. Retaining only the leading contributions for $t_m \to 0$, we get
\begin{align}
\mathcal{U}_{\beta}^{m}(x,t_m) \simeq & ~\delta _{m,0} ~\delta_{\beta,1}~\gamma ~\theta \left(vt_m-x \right), \\
\mathcal{W}_{m}(x,t_m) \simeq &~\delta _{m,0}~\theta \left(vt_m-x \right).
\end{align} 
\begin{figure}[t]
	\includegraphics[scale=1.15]{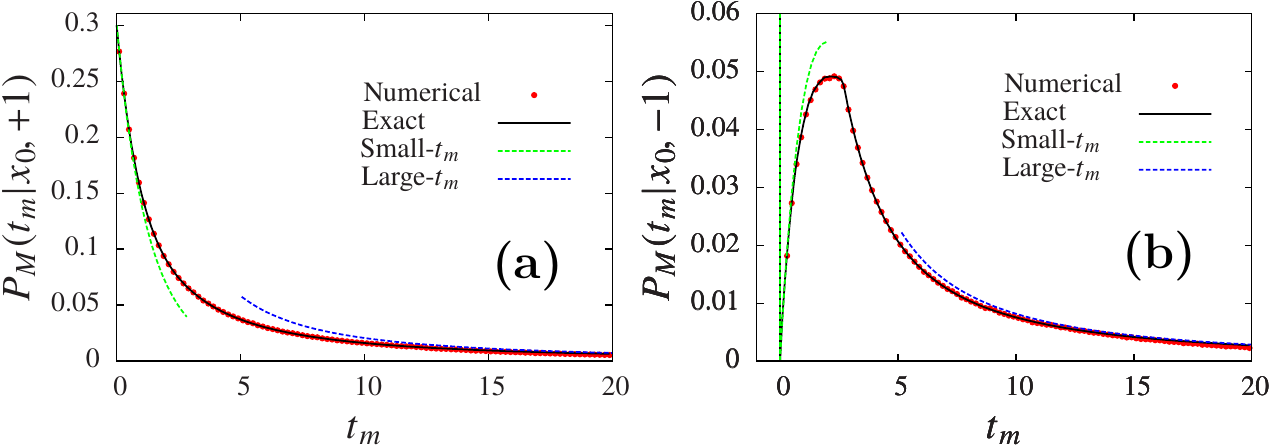}
	\centering
	\caption{Analytical forms of the distribution $P_{M}(t_m|x_0, {\pm 1})$ in Eqs. \eqref{Ptm-p} and \eqref{Ptm-n} have been plotted and compared with the numerical simulations for $x_0=2$, $\gamma =0.5$ and $v=1.5$. In addition, we have also plotted the asymptotic expressions of $P_{M}(t_m|x_0, {\pm 1})$ given in Eqs. \eqref{small-ptm} and \eqref{large-ptm}.}    
	\label{PtM-pic-2}
\end{figure}
Plugging these forms in Eqs. \eqref{Ptm-p} and \eqref{Ptm-n}, one can easily carry out the integration over $M$ and obtain the following approximate forms of distributions $P_{M}(t_m|x_0, \pm 1)$ for $t_m \to 0$:
\begin{align}
\begin{rcases}
&P_{M}(t_m|x_0, +1) ~\simeq~ \frac{v ~e^{-\gamma t_m}}{x_0 + vt_m + \frac{v}{\gamma}},  \\
& P_{M}(t_m|x_0, -1) ~\simeq~ \frac{v}{\gamma x_0 +v} ~\delta(t_m) + \frac{v \gamma t_m e^{-\gamma t_m}}{2 \left(x_0+\frac{v}{\gamma} \right)},~~~~~~~~
\end{rcases}
~~~~~~~\text{as }t_m \to 0.
\label{small-ptm}
\end{align}
On the other hand, for large $t_m$, we show in \ref{appen-large-Ptm} that the distributions $P_{M}(t_m|x_0, {\pm 1})$ take form
\begin{align}
\begin{rcases}
~~~~~~~~~~~~~~~~& P_{M}(t_m|x_0, +1) ~\approx~ \frac{\log2}{\sqrt{4 \pi D t_m^3}} \left( x_0 +\frac{v}{\gamma} \right), ~~~~~~~~\\
&P_{M}(t_m|x_0, -1) ~\approx~ \frac{x_0\log2}{\sqrt{4 \pi D t_m^3}}, 
\end{rcases}
~~~~~~~\text{as }t_m \to  \infty,
\label{large-ptm}
\end{align}
with $D = v^2 / 2 \gamma$. Recall that for Brownian motion, $P_{BM}(t_m|x_0) ~\approx~ \frac{\sqrt{D}}{x_0 \sqrt{\pi t_m}}$ as $t_m \to 0$ and $P_{BM}(t_m|x_0)  ~\approx~ \frac{x_0\log2}{\sqrt{4 \pi D t_m^3}} $ as $t_m \to \infty$ \cite{tmax-1}. Thus, the distribution has power-law form at both large and small tails. On the other hand, the small-$t_m$ behaviour for the active case is very different. As seen in Eq. \eqref{small-ptm}, the distribution goes to a constant value as $t_m \to 0$ when the particle starts from $x_0$ with a positive initial velocity. Similarly, for the negative initial velocity, it takes the form of a delta-function $[\sim \delta(t_m)]$ as $t_m \to 0$. Recall that this form arises due to the fact that the RTP, starting from $x_0$ with $-v$, has a non-zero probability to reach the origin without crossing $x_0$ again. Then, the maximum displacement is essentially $M=x_0$ and it is reached at time $t_m = 0$.

Coming to the large-$t_m$ behaviour in Eq. \eqref{large-ptm}, we see that for RTP also, the distributions $P_{M}(t_m|x_0, \pm 1)$ have power-law decay of the form $\sim t_m^{-{3/2}}$ just like the passive Brownian case. But there is one crucial difference. While for Brownian motion, the prefactor associated with this power-law decay vanishes exactly at $x_0 = 0$, it is non-zero for the active RTP case when it starts from the position of the absorbing wall with positive velocity. This means that at large $t_m$, the RTP with negative initial velocity behaves like a Brownian particle with effective diffusion constant $D = v^2 / 2 \gamma$ and also effective initial position $\left( x_0 + v /\gamma \right)$. The signature of activity is seen in the prefactor that characterises the power-law decay. Such observations have also been made for the survival problems of a RTP {\cite{FPT-7}}. In Fig. \ref{PtM-pic-2}, we have numerically verified the asymptotic expressions of $P_{M}(t_m|x_0, {\pm 1})$ in Eqs. \eqref{small-ptm} and \eqref{large-ptm}.

\section{Conclusion}
\label{conclusion}
In the first part of this paper, we have analytically computed the exact time-dependent position distributions of a one dimensional RTP in a finite interval $[0,M]$ with absorbing conditions at its two ends. Our study reveals rather interesting features of these distributions. For instance, the probability of finding the particle near the absorbing walls does not vanish  as seen in Figs. \ref{probpic1} and \ref{probpic2}. Similarly due to the persistent nature of the RTPs, the particle has non-vanishing distribution even when the initial position coincides with the positions of the walls, i.e.  $x_ 0 \to 0^+$ or $x_0 \to M^{-}$. Such peculiar behaviours are not observed for the passive Brownian motion \cite{Redner}. Another interesting observation in Figs. \ref{probpic1}(b) and \ref{probpic2}(b) is that the distributions $P_{\pm}(x,t|x_0, \mp 1)$ possess discontinuities at some position (say $x^{\ast}$) for some values of $t$. We provided a physical reasoning of these discontinuities based on the flipping ability of the RTPs from the boundary. 

In the second part of our paper, we investigated the statistics of the maximum displacement $M$ attained by the RTP before its first passage to the origin and the time $t_m$ at which this maximum value is achieved. Based on a path decomposition technique for Markov processes {\cite{tmax-2}, we derive a formula in Eq. \eqref{new-JT-1} that connects the joint distribution of $M$ and $t_m$ with the position distributions derived in the first part. Using this key connection, we obtain analytically the exact form of the joint distribution $\mathscr{P}(M,t_m|x_0, \sigma _i)$, where $\sigma _i \in \{ +1,-1 \}$ is the direction of the initial velocity. Next we use Eq. \eqref{new-JT-1} and maginalize it to obtain the distributions of $M$ and $t_m$. These expressions are provided in Eqs. \eqref{PPmargM-eq-3} and \eqref{PPmargM-eq-4} for $M$ and in Eqs. \eqref{Ptm-p} and \eqref{Ptm-n} for $t_m$. Following this exact analysis, we showed that the behaviour of the distribution $P_M(t_m|x_0, \sigma _i)$ of $t_m$ for smaller values of $t_m$ is completely different than that of the Brownian motion. Recall that for Brownian motion, this distribution has power-law tails both: $P_M^{BM}(t_m|x_0) \sim t_m ^{-1/2}$ for small $t_m$ and $P_M^{BM}(t_m|x_0) \sim x_0 ~t_m ^{-3/2}$ for large $t_m$ {\cite{tmax-1}}. Contrarily, for RTP, we find that $P_M(t_m|x_0, \sigma _i) ~\simeq~ \gamma v / (\gamma x_0 + v)$ goes to a constant value as $t_m \to 0$ for positive $\sigma _i$ whereas $P_M(t_m|x_0, \sigma _i) \sim \delta (t_m)$ has a delta function peak for negative $\sigma _i$. These forms are given in Eq. \eqref{small-ptm}.
For large $t_m$, on the other hand, the distribution $P_M(t_m|x_0, \sigma _i)$ for RTP also has a power-law decay of the form $\sim t_m ^{-3/2}$ which is similar to the Brownian motion result. But there is one crucial difference as shown in Eq. \eqref{large-ptm}. While for negative $\sigma _i$, the amplitude associated with the power-law grows linearly with the initial position as $\sim x_0$, it varies as $\sim (x_0 + v / \gamma)$ for the positive $\sigma _i$ case. This means that in contrast to the passive case, the amplitude of the power-law decay, for RTP (with positive $\sigma _i$), does not vanish even when $x_0 \to 0$, i.e. if the particle starts from the absorbing wall. Therefore the distribution of $t_m$ retains the signature of the active dynamics even at large $t_m$. Analogous behaviour has also been seen for the persistent properties of the RTP {\cite{FPT-7}}.

Our work can be extended to other directions. For example, it would be interesting to see how these distributions get affected for RTP with non-instantaneous tumbles. Another direction would be to study these quantities for RTPs in inhomogeneous media which is a more realistic model of active motion performed by certain bacteria \cite{Berg2003}. Finally it would also be interesting to extend these studies to other models of active particles.

\section*{Acknowledgement}
The authors acknowledge the support of the Department of Atomic Energy, Government of India, under project no.19P1112R\&D. A.K. acknowledges the support of the core research grant  CRG/2021/002455 and MATRICS grant MTR/2021/000350 from the Science and Engineering Research Board (SERB), Department of Science and Technology, Government of India.

\appendix
\section{Derivation of $\bar{\mathcal{P}}_{\pm}(x,s|x_0, +1)$}
\label{propagator_abs}
In this appendix, we will solve Eqs. \eqref{LT-FP-2} to obtain the Laplace transforms $\bar{\mathcal{P}}_{\pm}(x,s|x_0, +1)$ written in Eqs. \eqref{prob-LT-1} and \eqref{prob-LT-2}. To this end, we consider Eqs. \eqref{LT-FP-2} for $x \neq x_0$ for which we can drop the delta-function term. This yields
\begin{align}
\left[~v \partial_x+s~\right]\bar{\mathcal{P}}_{+}(x,s|x_0, +1)&=\gamma \bar{\mathcal{P}}_-(x,s|x_0, +1) ,\label{abs_appen2} \\
\left[-v \partial_x+s\right]\bar{\mathcal{P}}_{-}(x,s|x_0, +1)&=\gamma \bar{\mathcal{P}}_+(x,s|x_0, +1).
\label{abs_appen3}
\end{align}
Operating both sides of Eq. \eqref{abs_appen2} by $\left(- v \partial_x+s\right)$, we get
\begin{align}
\left[-v \partial_x+s\right] \left[~v \partial_x+s~\right]\bar{\mathcal{P}}_{+}(x,s|x_0, +1)&=\gamma^2 \bar{\mathcal{P}}_+(x,s|x_0, +1). \label{abs_appen4} 
\end{align}
We next try the solution $\bar{\mathcal{P}}_{+}(x,s|x_0, +1) \sim e^{\frac{\omega x}{v}} $ and plug it in Eq. \eqref{abs_appen4}. One then gets $\omega=\pm \lambda (s)$ where $\lambda(s)=\sqrt{s^2-\gamma^2}$. Then the solution for $\bar{\mathcal{P}}_{+}(x,s|x_0, +1)$ is given by
\begin{align}
\bar{\mathcal{P}}_+(x,s|x_0, +1)=
    \begin{cases}
    A_1(s) e^{\frac{\lambda(s) x}{v}}+A_2(s) e^{-\frac{\lambda(s) x}{v}}, & \text{if}\ 0<x<x_0~~~~~~~~~~~~~~~~~~~~~~~~~ \\
     B_1(s) e^{\frac{\lambda(s) x}{v}}+B_2(s) e^{-\frac{\lambda(s) x}{v}}, & \text{if}\ x_0<x<M.
    \end{cases}
\label{abs_appen6}
\end{align}
Here $A_1(s),~ A_2(s),~ B_1(s)$ and $B_2(s)$ are functions of $s$ but do not depend on $x$. The other Laplace transform $\bar{\mathcal{P}}_{-}(x,s|x_0, +1) $ can be obtained by inserting the solution for $\bar{\mathcal{P}}_{+}(x,s|x_0, +1) $ in Eq. \eqref{abs_appen2}. The expression reads
\begin{align}
\bar{\mathcal{P}}_-(x,s|x_0, +1)=
    \begin{cases}
    A_1(s) e^{\frac{\lambda(s) x}{v}}\left[\lambda(s)+s\right]+A_2(s) e^{-\frac{\lambda(s) x}{v}}\left[\lambda(s)-s\right], & \text{if}\ 0<x<x_0 \\
     B_1(s) e^{\frac{\lambda(s) x}{v}}\left[\lambda(s)+s\right]+B_2(s) e^{-\frac{\lambda(s) x}{v}}\left[\lambda(s)-s\right], & \text{if}\ x_0<x<M.
    \end{cases}
\label{abs_appen7}
\end{align}
The task now is to compute the functions $A_1(s),~ A_2(s),~ B_1(s)$ and $B_2(s)$ which appear in the expressions of $\bar{\mathcal{P}}_{\pm}(x,s|x_0, +1)$. For this, we integrate both sides of Eqs. \eqref{LT-FP-2} from $-\epsilon$ to $\epsilon$ and take $\epsilon \to 0^+$ limit. This gives rise to the following two conditions in $\bar{\mathcal{P}}_{\pm}(x,s|x_0, +1) $:
\begin{align} 
  \bar{\mathcal{P}}_+(x\to x_0^+,s|x_0, +1)-  \bar{\mathcal{P}}_+(x\to x_0^-,s|x_0, +1) &=\frac{1}{v},   \label{abs_appen8} \\
  \bar{\mathcal{P}}_-(x\to x_0^+,s|x_0, +1)-  \bar{\mathcal{P}}_-(x\to x_0^-,s|x_0, +1) &=0.     
\label{abs_appen9} 
\end{align}
The other two conditions follow from the behaviour of $\bar{\mathcal{P}}_{\pm}(x,s|x_0, +1) $ near the absorbing boundaries. These conditions are written in Eqs. \eqref{bc1} and \eqref{bc2} which can be suitably transformed in the Laplace domain as $\bar{\mathcal{P}}_+(x \to 0,s|x_0, +1)=0$ and $\bar{\mathcal{P}}_-(x \to M,s|x_0, +1)=0$. Inserting the expressions of $\bar{\mathcal{P}}_{\pm}(x,s|x_0, +1) $ from Eqs. \eqref{abs_appen6} and \eqref{abs_appen7} into these conditions, we find 
\begin{align} 
\begin{split}
  A_1(s)&=-A_2(s)=\frac{s-\lambda(s)}{2 v \lambda(s)} \Big[ \mathcal{Z}(x_0,s)- \mathcal{Z}(2 M-x_0,s)\Big],\\
  B_1(s)&=-\mathcal{F}(s)\frac{s+\lambda(s)}{2 v \lambda(s)} \Big[ \mathcal{Z}(2 M-x_0,s)-\mathcal{F}(s) \mathcal{Z}(2 M+x_0,s)\Big],\\
   B_2(s)&=\frac{s+\lambda(s)}{2 v \lambda(s)} \Big[ \mathcal{Z}(-x_0,s)-\mathcal{F}(s) \mathcal{Z}(x_0,s)\Big] ,
   \end{split}
   \label{abs_appen99}
\end{align}
where the functions $\mathcal{F}(s)$ and $\mathcal{Z}(x,s)$ are defined as 
\begin{align} 
 \mathcal{F}(s)=\frac{s-\lambda(s)}{s+\lambda(s)},~~~~~~ \mathcal{Z}(x,s)=\frac{e^{-\frac{\lambda(s) }{v}x}}{1-\mathcal{F}(s) e^{-\frac{2 \lambda(s)M}{v}}}.
\label{abs_appen10}    
\end{align}
Using Eq. \eqref{abs_appen99} in Eqs. \eqref{abs_appen6} and \eqref{abs_appen7}, we get exact form of the Laplace transforms $\bar{\mathcal{P}}_{\pm}(x,s|x_0, +1) $ as
\begin{align}
\bar{\mathcal{P}}_+(x,s|x_0,+1)=&
    \frac{s+\beta \lambda(s)}{2 v \lambda(s)}\Big[ \left\{ \mathcal{F}(s) \right\}^{1+\beta}\mathcal{Z}(2 M-|x_0-x|,s)-\left\{ \mathcal{F}(s) \right\}^{\frac{1+\beta}{2}}\mathcal{Z}(2 M-x_0-x,s) \nonumber \\
 & ~~~~~~~~~~~+\mathcal{Z}(|x_0-x|,s)-\left\{ \mathcal{F}(s) \right\}^{\frac{1+\beta}{2}}\mathcal{Z}(x_0+x,s) \Big],\\
  \bar{\mathcal{P}}_-(x,s|x_0,+1)=&\frac{\gamma}{2 v \lambda(s)}\Big[\mathcal{F}(s) \mathcal{Z}(2 M-|x_0-x|,s)-\mathcal{Z}(2 M-x_0-x,s) \nonumber \\
  & ~~~~~~~~~~~+\mathcal{Z}(|x_0-x|,s)-\mathcal{F}(s)\mathcal{Z}(x_0+x,s) \Big],
\end{align}
where $\beta = \text{sgn}(x-x_0)$. These expressions have been quoted in Eqs. \eqref{prob-LT-1} and \eqref{prob-LT-2} in the main text.

\section{Asymptotic expressions of $P_{M}(t_m|x_0, {\pm 1})$ for large $t_m$}
\label{appen-large-Ptm}
Here, we derive the large-$t_m$ behaviour of the distributions $P_{M}(t_m|x_0, {\pm})$ which was written in Eq. \eqref{large-ptm} in the main text. For simplicity, we consider this calculation separately for $P_{M}(t_m|x_0, -1)$ and $P_{M}(t_m|x_0, +1)$.
\subsection{$P_{M}(t_m|x_0, -1)$ as $t_m \to \infty$}
To begin with, we look at the joint distribution $\mathscr{P}(M,t_m|x_0, -)$ of $M$ and $t_m$ in Eq. \eqref{new-JT-3}. For large $t_m$, we can drop the $\delta(t_m)$ term in this expression and rewrite it as
\begin{align}
& \mathscr{P}(M,t_m|x_0, -1) \simeq  \gamma P_{+}(M,t_m|x_0, -1)E_-(M,M). \label{appen-large-eq-1}
\end{align}
Inserting the forms of $P_{+}(M,t_m|x_0, -1)$ and $E_-(M,M)$ from Eqs. \eqref{pp2} and \eqref{ee1}, we obtain
\begin{align}
\mathscr{P}(M,t_m|x_0, -1) & \simeq  \frac{\gamma e^{-\gamma t_m}}{2 \left(M +\frac{v}{\gamma} \right)} \left[ \mathcal{W}_{0}(M-x_0,t_m)-\mathcal{W}_{1}(M-x_0,t_m) \right. \nonumber \\
&~~~~~~~~ \left. -~\mathcal{W}_{0}(M+x_0,t_m) +\mathcal{W}_{1}(M+x_0,t_m) \right], \label{appen-large-eq-2}
\end{align}
where $\mathcal{W}$-functions are given in Eq. \eqref{W-fun}. To simplify Eq. \eqref{appen-large-eq-2} further, we take the Laplace transformation with respect to $t_m~(\to s)$. Then the large-$t_m$ behaviour of $\mathscr{P}(M,t_m|x_0, -1)$ in the time domain will correspond to the small-$s$ behaviour of the Laplace transform $\bar{\mathscr{P}}(M,s|x_0, -1)$. Now to write $\bar{\mathscr{P}}(M,s|x_0, -1)$ from Eq. \eqref{appen-large-eq-2}, one needs to take the Laplace transformation of various $\mathcal{W}$-functions. Fortunately this can be done and is written in Eq.  \eqref{abs_appen13}. We then obtain
\begin{align}
\bar{\mathcal{P}}_{max}(M,s|x_0, -1) \simeq \frac{\gamma }{ \left( M + \frac{v}{\gamma}\right)} \left[ \frac{\text{exp} \left(-\frac{\lambda(s+\gamma)~(M-x_0)}{v} \right)-\text{exp} \left(-\frac{\lambda(s+\gamma)~(M+x_0)}{v} \right)}{\{  \gamma +s + \lambda (s+\gamma) \}   \left\{1-\mathcal{F}(s+\gamma)~\text{exp} \left(-\frac{2 \lambda(s+\gamma)M}{v} \right) \right\} } \right],
 \label{appen-large-eq-3}
\end{align}
where $\lambda(s) = \sqrt{s^2 - \gamma ^2}$ and $\mathcal{F}(s)$ is given in Eq. \eqref{abs_appen10}. For $s \to 0$, one has $\lambda(s+\gamma) \simeq \sqrt{2 \gamma s}$ and $\mathcal{F}(s + \gamma) \simeq 1$ and using these approximations in Eq. \eqref{appen-large-eq-3}, we find
\begin{align}
\bar{\mathscr{P}}(M,s|x_0, -1) \simeq  \frac{1}{M+\frac{v}{\gamma}} \left[ \frac{e^{-\sqrt{\frac{s}{D}}(M-x_0)} -e^{-\sqrt{\frac{s}{D}}(M+x_0)} }{1-e^{-2\sqrt{\frac{s}{D}}M}} \right],~~~~\text{as }s \to 0,  \label{appen-large-eq-4}
\end{align}
with $D = v^2 / 2 \gamma $. Next, we use Eq. \eqref{Laplace_appen01081} to perform the inverse Laplace transformation of $\bar{\mathscr{P}}(M,s|x_0, -1)$ as
\begin{align}
\mathscr{P}(M,t_m|x_0, -1) \simeq \frac{2 \pi D}{M^2 \left(M+\frac{v}{\gamma} \right)}~\sum _{n=1}^{\infty} (-1)^{n+1} n  \sin \left( \frac{n \pi x_0}{M} \right)e^{-\frac{n^2 \pi ^2 D t_m}{M^2}}.
\end{align}
Finally integrating this joint distribution over $M$, we obtain the marginal distribution $P_{M}(t_m|x_0, -1)$ as
\begin{align}
P_{M}(t_m|x_0, -1) \simeq 2 \pi D \sum _{n=1}^{\infty} (-1)^{n+1} n \int _{x_0}^{\infty} \frac{dM}{M^2 \left(M+\frac{v}{\gamma} \right)}~\sin \left( \frac{n \pi x_0}{M} \right)e^{-\frac{n^2 \pi ^2 D t_m}{M^2}}. \label{appen-large-eq-5}
\end{align}
Changing the variable $M = n \pi \sqrt{D t_m} ~y $ and taking $t_m \to \infty $ with $y$ fixed, we can simlify Eq. \eqref{appen-large-eq-5} to get
\begin{align}
P_{M}(t_m|x_0, -1) \simeq = \frac{2 x_0}{\pi \sqrt{D t_m^3}}~\sum _{n=1}^{\infty} \frac{(-1)^{n+1}}{n}~\int_{0}^{\infty} \frac{dy}{y^4}~\text{exp}\left( -\frac{1}{y^2} \right).
\end{align}
Both the summation and the integration can now be carried out explicitly as
\begin{align}
\sum _{n=1}^{\infty} \frac{(-1)^{n+1}}{n} = \log 2,~~~\text{and}~~~\int_{0}^{\infty} \frac{dy}{y^4}~\text{exp}\left( -\frac{1}{y^2} \right) = \frac{\sqrt{\pi}}{4},
\end{align}
which then gives
\begin{align}
P_{M}(t_m|x_0, -1) \simeq \frac{ x_0 \log 2}{ 2\sqrt{\pi D t_m^3}},~~~~~~\text{as }t_m \to \infty. \label{appen-large-eq-6}
\end{align}
\subsection{$P_{M}(t_m|x_0, +1)$ as $t_m \to \infty$}
We next look at the asymptotic behaviour of $P_{M}(t_m|x_0, +1)$ as $t_m \to \infty$. Once again, we begin with the joint distribution $\mathscr{P}(M,t_m|x_0, +1)$ in Eq. \eqref{new-JT-2} and rewrite it in the large $t_m$ limit as
\begin{align}
\mathscr{P}(M,t_m|x_0, +1) & \simeq \frac{e^{-\gamma t_m}}{2 ~\left( M + \frac{v}{\gamma} \right)}  \left[2 \delta \left( t_m - \frac{M-x_0}{v} \right)+ \mathcal{U}_{1}^{0}(M-x_0,t_m) \right. \nonumber \\
&~~~ \left.-\mathcal{U}_{1}^{1}(M-x_0,t_m) -~\mathcal{U}_{1}^{1}(M+x_0,t_m)+\mathcal{U}_{1}^{2}(M+x_0,t_m) \right],\label{appen-large-eq-7}
\end{align}
where the $\mathcal{U}$-functions are given in Eq. \eqref{U-fun}. As before, we proceed to take the Laplace transformation of $\mathscr{P}(M,t_m|x_0, +1)$ in Eq. \eqref{appen-large-eq-7} with respect to $t_m$. For this, one needs the Laplace transformations of different $\mathcal{U}$-functions which have been specified in Eq. \eqref{abs_appen12}. Using this equation, we get
\begin{align}
\bar{\mathscr{P}}(M,s|x_0, +1) \simeq  \frac{\gamma \left[  \{  \gamma +s + \lambda (s+\gamma) \} e^{-\frac{\lambda(s+\gamma)~(M-x_0)}{v} }- \{  \gamma +s - \lambda (s+\gamma) \}  ~e^ {-\frac{\lambda(s+\gamma)~(M+x_0)}{v} }   \right] }{ \left( M + \frac{v}{\gamma}\right) \{  \gamma +s + \lambda (s+\gamma) \}    \left\{      1-\mathcal{F}(s+\gamma)~e^{-\frac{2 \lambda(s+\gamma)M}{v} }    \right\}  }. \label{appen-large-eq-8}
\end{align}
For small values of $s$, we again use the approximations $\lambda(s+\gamma) \simeq \sqrt{2 \gamma s}$ and $\mathcal{F}(s + \gamma) \simeq 1$. We also use $ \left( \gamma +s \pm \lambda (s+\gamma) \right) \simeq \left( \gamma \pm \sqrt{2 \gamma s} \right) \simeq \gamma e^{ \pm \sqrt{\frac{2s}{\gamma}} }$. Then the Laplace transform $\bar{\mathscr{P}}(M,s|x_0, +1) $ in Eq. \eqref{appen-large-eq-8} takes the form
\begin{align}
\bar{\mathscr{P}}(M,s|x_0, +1) \simeq  \frac{1}{M+\frac{v}{\gamma}} \left[ \frac{e^{-\sqrt{\frac{s}{D}}\left(M-x_0 - \frac{v}{\gamma} \right)} -e^{-\sqrt{\frac{s}{D}}\left( M+x_0 +\frac{v}{\gamma}\right)} }{1-e^{-2\sqrt{\frac{s}{D}}M}} \right],~~~~\text{as }s \to 0.  \label{appen-large-eq-49}
\end{align}
This has exactly the same form as $\bar{\mathscr{P}}(M,s|x_0, -1)$ in Eq. \eqref{appen-large-eq-4} but with $x_0$ replaced by $\left( x_0 + v / \gamma \right)$. Repeating exactly the same analysis as for $\bar{\mathscr{P}}(M,s|x_0, -1)$, we obtain 
\begin{align}
P_{M}(t_m|x_0, +1) \simeq \frac{\log2}{\sqrt{4 \pi D t_m^3}} \left( x_0 +\frac{v}{\gamma} \right).
\end{align}

\section{Some useful Laplace transfroms}
\label{Laplace}
In this appendix, we provide a list of some inverse Laplace transformations which will be useful in deriving various results in the paper. More details can be found in \cite{laplace_table}. For a function $f(t)$, the Laplace transformation is given by $\bar{f}(s) = \int _0^{\infty}dt e^{-st} f(t)$ and the inverse Laplace transformation is denoted by $L_{s} \left[ \bar{f}(s) \right] = f(t)$. 
\begin{align}
&L_s \left[\frac{s+\gamma}{\lambda (s)} \mathcal{F}^m \mathcal{Z}(x,s) \right]=\sum_{n=0}^{\infty}\frac{\gamma}{2}\left[g_{2n+2m-1}(x+2nM,t)+2 g_{2n+2m}(x+2nM,t)\right. \nonumber \\
&~~~~~~~~~~~~~~~~~~~~~~~~~~~~~~~~~\left.+g_{2n+2m+1}(x+2nM,t)\right],~~~~~~~~~~~~\text{where } m\geq1.\label{Laplace_appen1}\\
&L_s \left[\frac{s+\gamma}{\lambda (s)} \mathcal{Z}(x,s) \right]=\delta (t-\frac{x}{v})+\sum_{n=0}^{\infty}\frac{\gamma}{2}\left[g_{2n+1}(x+2(n+1)M,t)+2 g_{2n}(x+2nM,t)\right. \nonumber \\
&~~~~~~~~~~~~~~~~~~~~~~~~~~~~~~~\left.+g_{2n+3}(x+2(n+1)M,t)\right]+\frac{\gamma t}{t-\frac{x}{v}}g_1(x,t).\label{Laplace_appen2}\\
&L_s \left[\mathcal{F}^m \mathcal{Z}(x,s) \right]=-v\frac{d}{dx}\Big[\sum_{n=0}^{\infty}g_{2n+2m}(x+2nM,t)\Big],~~~~~~ \text{ where }m\geq0.\label{Laplace_appen4}\\
\label{Laplace_appen5}\\
& L_s\left[ \frac{e^{-\sqrt{\frac{s}{D}}(M-x_0)} -e^{-\sqrt{\frac{s}{D}}(M+x_0)} }{1-e^{-2\sqrt{\frac{s}{D}}M}} \right] = \frac{2 D \pi}{M^2} \sum _{n=0}^{\infty} (-1)^{n+1} n  \sin \left( \frac{n \pi x_0}{M} \right)e^{-\frac{n^2 \pi ^2 D t}{M^2}}. \label{Laplace_appen01081}    \\
&L_s \left[ \frac{1}{\sqrt{4 D s}} \frac{e^{-\sqrt{\frac{s}{D}}\mid x \mid }}{\left(1-e^{-2\sqrt{\frac{s}{D}}L }\right)}\right]=\sum_{n=0}^{\infty}\frac{1}{ \sqrt{4\pi Dt}}e^{-\frac{(\mid x \mid +2nM)^2}{4 D t}}. \label{Laplace_appen01}\\
&L_s\left[ \frac{s+\beta \lambda(s)}{\lambda(s)} \mathcal{F}^{m} \mathcal{Z}(x,s)\right]=2 \delta _{m,0} ~\delta_{\beta,1}~ \delta \left(t-\frac{x}{v} \right)+\mathcal{U}_{\beta}^{(m)}(x,t), \text{ with } \beta = \pm 1, m\geq0. \label{abs_appen12}\\
&L_s \left[\frac{\mathcal{F}^m}{\lambda (s)}  \mathcal{Z}(x,s) \right]=\mathcal{W}_m(x,t),~~~~~~~~~~~\text{ where }m\geq0.
\label{abs_appen13}
\end{align}
In these inverse Laplace transforms, the functions $\mathcal{U}_{\beta}^{(m)}(x,t)$, $\mathcal{W}_m(x,t)$ and $g_m(x,t)$ are given in Eqs. \eqref{U-fun-2}-\eqref{g-fun}, $\mathcal{Z}(x,s)$ and $\mathcal{F}(s)$ are given in Eq. \eqref{abs_appen10} and $\lambda(s) = \sqrt{s^2 -\gamma ^2}$.

\end{document}